\documentclass[aps,a4paper,twocolumn,twoside]{revtex4}

\usepackage{epsfig}
\usepackage{epstopdf}
\usepackage{amsmath,amssymb,color,amsthm}
\usepackage{tikz}
\usepackage{color}

\usepackage[english]{babel}

\parskip=\medskipamount

\newcommand{\eq}[1]{(\ref{#1})}
\newcommand{\fig}[1]{Fig.~\ref{#1}}

\newcommand{\be}{\begin{equation}}
\newcommand{\ee}{\end{equation}}

\newcommand{\barr}{\begin{array}}
\newcommand{\earr}{\end{array}}

\newcommand{\beqn}{\begin{eqnarray}}
\newcommand{\eeqn}{\end{eqnarray}}

\newcommand{\bw}{\begin{widetext}}
\newcommand{\ew}{\end{widetext}}

\newcommand\disp{\displaystyle}

\newcommand{\eps}{\varepsilon}
\newcommand{\la}{\left<}
\newcommand{\ra}{\right>}

\def\runninghead#1#2{\pagestyle{myheadings}
\markboth{{\protect\it{\quad #1}}\hfill} {\hfill{\protect\it{#2\quad}}}}

\begin{document}

\runninghead{\sl S.K. Nechaev, A. Sobolevski, O.V. Valba}{\sl Planar diagrams from optimization}

\title{Planar diagrams from optimization}

\author{S.K. Nechaev$^{1,2}$, A.N. Sobolevski$^{3,4}$, O.V. Valba$^{1,5}$}

\affiliation{$^{1}$LPTMS, Universit\'e Paris Sud, 91405 Orsay Cedex,
  France \\
$^{2}$P.N. Lebedev Physical Institute of the Russian Academy of
Sciences, 119991, Moscow, Russia \\
$^{3}$Institute for Information Transmission Problems of the Russian
Academy of Sciences (Kharkevich Institute), 127994 Moscow, Russia \\
$^{4}$National Research University Higher School of Economics, 101000 Moscow, Russia \\
$^{5}$Moscow Institute of Physics and Technology, 141700, Dolgoprudny, Russia}

\date{\today}

\begin{abstract}

We propose a new toy model of a heteropolymer chain capable of forming planar secondary structures
typical for RNA molecules. In this model the sequential intervals between neighboring monomers along a
chain are considered as quenched random variables. Using the optimization procedure for a
special class of concave--type potentials, borrowed from optimal transport analysis, we derive the
{\em local} difference equation for the ground state free energy of the chain with the planar
(RNA--like) architecture of paired links. We consider various distribution functions of intervals
between neighboring monomers (truncated Gaussian and scale--free) and demonstrate the existence of
a topological crossover from sequential to essentially embedded (nested) configurations of paired
links.

\end{abstract}

\maketitle

\section{Introduction}

Both DNAs and RNAs are heteropolymers constituting of four different nucleotide types. The
peculiarity of RNA chains consists in the additional freedom of the formation of complex secondary
structures. These secondary (intra--molecular) structures are stabilized by theromoreversible
hydrogen bonds between non--neighboring nucleotides and mostly take a ``cactus--like''
hierarchically folded form, topologically isomorphic to a tree. The structures which do not belong
to this tree--like class are known in the literature as ``pseudoknots'', and in most cases are
highly suppressed. The main task of any computational algorithm predicting the secondary structure
of RNA can be formulated as a search for a secondary structure with the lowest value of the free
energy (``ground state'') among all allowed cactus--like structures.

\begin{figure}[ht]
\epsfig{file=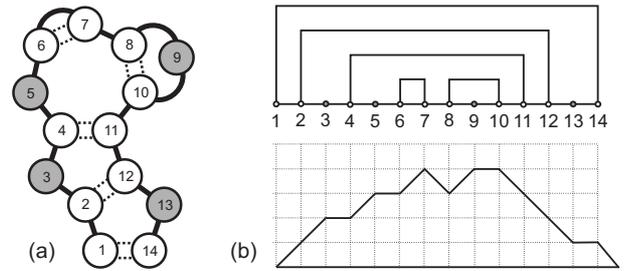, width=8cm}
\caption{(a) Schematic cactus--like secondary structures of an RNA--like chain; (b) the height
diagram for (a) represented by a Motzkin path.}
\label{fig:1}
\end{figure}

Construction of an effective dynamic programming algorithm (DPA) to predict RNA--like secondary
structures is a much more challenging problem than that for a classical DNA--matching problem (see
\cite{W,WV,HWA2,Hwa97,W2}). In the simplest possible case the generic DPA allowing to calculate the
cost function and to find the ground state structure of an RNA--type polymer with a given primary
sequence, is as follows. Suppose that a given chain consists of $n$ monomer units, each unit chosen
from a set of $c$ different types (letters) A, B, C, D, \dots\,. These units can form noncovalent
bonds with each other, at most one bond per unit. The energy of a bond depends on which letters are
bonded, the simplest choice is to assign some attraction energy $u$ to the bonds between similar
letters (A--A, B--B, C--C, \dots) and zero energy the bonds between different letters (A--B, A--D,
B--D, \dots). In real RNAs matches are the interactions between {\em complimentary} nucleotides
rather than {\em similar} ones, which gives rise to a slightly different matrix of interactions.
However, at least for random RNAs this difference is irrelevant: it is important that the fraction
of possible matches is $\frac{1}{c}$, the rest corresponding to mismatches. Schematically the
secondary structure of RNA chain is shown in \fig{fig:1}a.

The simple model, serving as a ``shooting range'' for theoretical consideration of secondary
structures formation typical for the ensemble of messenger RNAs, is as follows. Let us neglect the
contribution of loop factors to the partition function and variation in the energies of different
types of complementary nucleotides, avoid the constraints on the minimal size of loops in the
structure, and disregard the stacking interactions (the cooperativity in bonds creation between
adjacent pairs of monomers). What is preserved only, is the possibility of a formation of a
cactus--like folded configurations for any arbitrary sequence of nucleotides. The partition
function of this model is known (see, for example, \cite{W2,bund,mueller,krz,hwa3,churchkhela}) to
satisfy the recursion relation:
\be
\left\{
\begin{array}{l}
\disp g_{i,i+k}=g_{i+1,i+k}+\sum_{s=i+1}^{i+k} \beta_{i,s}g_{i+1,s-1}\,
g_{s+1,i+k}; \medskip \\
\disp g_{i,i}=g_{i + 1,i}=1.
\end{array} \right.
\label{eq:1}
\ee
The term $g_{i,j}$ describes the contribution to the partition function of the part of the sequence
between monomers $i$ and $j$. The Boltzmann weights
\be
\beta_{i,j} = e^{-u_{i,j}/T},\quad 1\le i < j \le n
\label{eq:2}
\ee
are the statistical weights of bonds, and the ``boundary conditions'' $g_{i, i} = g_{i + 1, i} = 1$
take care of the unpaired bonds. Expression \eq{eq:1} is convenient for recursive computation. The
energy of the ground state, $F_{1,n}=\lim\limits_{T\to +0} -T\ln g_{1,n}$, is the free energy of
the system at zero temperature, so it can be calculated as follows:
\begin{multline}
F_{i,i+k} = \lim_{T \to +0} -T\ln g_{i,i+k} = \min \Big\{F_{i+1,i+k}, \\
\min_{s=i+1,\dots,i+k}\big[u_{i,s}+F_{i+1,s-1}+F_{s+1,i+k}\big]\Big\}.
\label{eq:3}
\end{multline}

The geometry of the secondary structure becomes very transparent if one represents binding of
monomers by so-called ``height diagram'' \cite{hwa3} depicted in the \fig{fig:1}b. That is,
construct an auxiliary one--dimensional walk according to a following rule. Start from $x=0$ and at
each discrete time tick allow a step of $\pm 1$, or $0$. If the monomer $i$ in the original
cactus--like structure is connected to a monomer $j$ and $i<j$, then $i$-th step of the walk is
``up.'' If $i$ is connected to such a $j$ that $i>j$, then the corresponding step is ``down.'' If
$i$ is not connected with any other monomers, then the walker at $i$-th step stays put. Clearly,
thus defined trajectory returns to zero after $n$ steps and remains non-negatve for all $0<i<n$,
i.e.\ stays in the domain $(x\ge 0, i\ge 1)$ on $(x,i)$--plane. Such trajectories, being discrete
Brownian excursions, are called Motzkin paths \cite{lando}. It is clear from the comparison of
\fig{fig:1}a and~b that there exists a one-to-one correspondence between cactus--like RNA
secondary structures and height diagrams represented by Mozkin paths. Namely, the height of the
point in the height diagram equals to the number of arcs going above the corresponding point on the
arc diagram, i.e.\ coincides with the number of bonds one has to break to reach the corresponding
monomer from the starting point of the chain. An important statistical characteristic of the state
of the system is the so-called ``roughness exponent,'' $\gamma$, which links the mean height, $\la
h \ra$, of such a diagram with the length, $L$, of the chain: $\la h\ra \sim L^{\gamma}$, $0
\leqslant \gamma \leqslant 1$.

For homopolymer RNAs, the interaction energies $u_{i,j}$ take one and the same value $u$
independent from $i$ and~$j$. It is well known that for the uniform model Eq.~\eq{eq:1} can be easily
solved exactly by generating functions method \cite{lando}. This model displays the existence of a
2nd order phase transition from unpaired to strongly paired regime at $u=u_{\rm cr}$. The roughness
exponent, $\gamma$, for a height diagram is typical for randomly branched homopolymer,
$\gamma=1/2$.

The investigation of thermodynamic properties of {\it random} RNA--like chains is addressed in a
number of recent theoretical papers \cite{krz,hwa3,orland1,orland2,weise}. In these works it has
been supposed that $u_{i,j}$ is a quenched uncorrelated random function of $i$ and $j$, having a
Gaussian distribution. Within such a model it has been demonstrated that the presence of a frozen
heteropolymer structure of a chain plays a crucial role: due to the frustrations in the primary
sequence, the system exhibits a glass transition \cite{hwa3,krz}. The quenched randomness in the
primary sequence affects also the height diagram. It was found numerically that in glassy state of
random RNA the roughness exponent $\gamma$ takes the value close to $\gamma=2/3$. Recent analytic
estimates by field--theoretic arguments and RG analysis \cite{weise} give $\gamma\simeq 5/8$.
Despite the essential progress in the field, to our point of view, the question about the value of
roughness exponent for random heteropolymer RNAs is still open.

\section{The random interval model: Subadditivity and Submodularity}

Let us begin with some general definitions relating topology of planar diagrams and optimization.
Following R. McCann \cite{McCann.R:1999}, we call the function $w$ a \emph{cost function of concave
type} if for any $x_1, x_2, y_1, y_2 \in \mathbb{R}$ the inequality
\be
w(x_1, y_1) + w(x_2, y_2) \le w(x_1, y_2) + w(x_2, y_1)
\label{eq:101}
\ee
implies that the intervals connecting $x_1$ to $y_1$ and $x_2$ to $y_2$ are either disjoint or one
of them is contained in the other. Examples are: $w(x, y) = |x - y|^{\alpha}$ with $0 < \alpha <
1$, or $w(x,y) = \ln |x - y|$ extended to the diagonal $x = y$ by $-\infty$. In fact, whenever a
cost function $w$ of concave type is spatially homogeneous and symmetric, i.e., $w(x, y) = g(|x -
y|)$, the function $g$ must be strictly increasing and strictly concave \cite{McCann.R:1999}. Let
now $x_1 < x_2 < \dots < x_{2n}$ be an even number of points on the real line $\mathbb{R}$.
Consider the complete graph $K_{2n}$ on these points, each of whose edges $(x_i, x_j)$ is equipped
with a weight $w(x_i, x_j)$. We look for a minimum--weight perfect matching in the graph $K_{2n}$,
i.e., for a set of $n$ edges such that the sum of their weights is minimal.

A bipartite version of the graph matching problem has been thoroughly treated for costs of concave
type in the continuous setting in \cite{McCann.R:1999}. Similar discrete versions have also been
considered in the literature on optimal algorithms construction for the specific case of the
distance $|x - y|$ \cite{Karp.R:1975, Werman.M:1986, Aggarwal.A:1992} and for a general cost
function of a concave type in \cite{Delon.J:2011a}. Call a matching
\emph{planar} if, for any two arcs $(x_i, y_i)$ and $(x_{j}, y_{j})$ that are present in the
matching, the corresponding intervals in $\mathbb{R}$ are either disjoint or one of them is
contained in the other. In \cite{Aggarwal.A:1992, McCann.R:1999} it is proved that a
minimum--weight matching is planar.

\begin{figure}[ht]
\epsfig{file=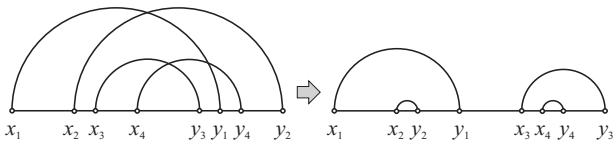, width=8cm}
\caption{Optimization with the concave--type cost function leads to the planar pairing.}
\label{fig:example}
\end{figure}

In \fig{fig:example} we rephrase this theorem pictorially. Taking weights $w(x_i,y_i)=
\ln|x_i-y_i|$, we can straightforwardly check that the minimal value of the total cost function
$\Omega(x_1,y_1;...; x_n,y_n)$, where
$$
\Omega(x_1,y_1;...; x_n,y_n)=\sum_{\rm \{arcs\}}\ln|x_i-y_i|,
$$
is achieved at some planar configuration of pairings.

\begin{figure}[ht]
\epsfig{file=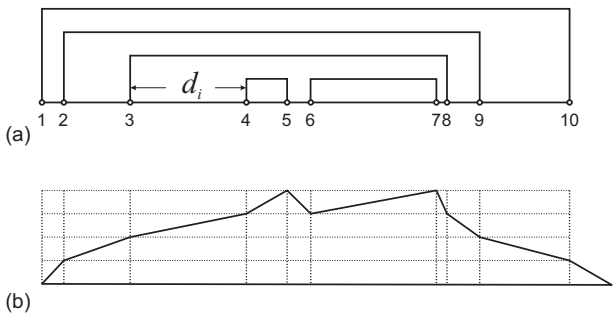, width=8cm}
\caption{Typical configuration of a random interval RNA, shown (a) by arcs, and (b) by a height
diagram.}
\label{fig:2}
\end{figure}

Now we are in position to formulate our toy Random Interval Model (RIM) of a quenched heteropolymer
RNA, in which the paired monomers interact with the energy $\eps_{i,j}$, which is a concave
function of the distance between monomers along the chain. In particular, we choose
$\eps_{i,j}$ of the form
\be
\eps_{i,j} = u \ln |x_i-x_j|; \qquad (j\neq i)
\label{eq:4}
\ee
where $u$ is some positive constant, and $x_i, x_j$ are the coordinates of monomers $i$ and $j$
along the chain. The distances $d_i=|x_{i+1}-x_i|$ along the chain between sequential monomers
capable to form pairs are quenched random variables taken independently from some distribution
$P(d_i=d)$. Schematically, a typical realization of a RIM is depicted in \fig{fig:2} by arcs (a)
and by a height diagram (b).

Let us emphasize that the key feature of the RIM consists in the fact that the interaction energy
between paired monomers, $\eps_{i,j}$, is a concave function of distance. In principle, one could
take $\eps_{i,j}$ in the form $\eps_{i,j}= u|x_j-x_i|^{\alpha_1}$, where $0<\alpha_1<1$, or
$\eps_{i,j}=-u|x_j-x_i|^{-\alpha_2}$, where $\alpha_2> 0$ ($j\neq i$). The main conclusions will
survive, though the details are model--dependent.

Supposing that every monomer in the ground state structure is involved in binding, after some
simplifications we get from \eq{eq:3}:
\be F_{i,i+k} =
\min_{s=i+1,i+3,\dots,i+k}\Big[\eps_{i,s}+F_{i+1,s-1}+F_{s+1,i+k}\Big]
\label{eq:5}
\ee
with the ``boundary conditions'' $F_{i + 1, i} = 0$ for any~$i$. Note that it is enough to extend
the $\min$ in $s$ over values with odd increments with respect to~$i$: no arc can cover an odd
number of points, because otherwise some of them would be excluded
from the structure due to planarity.

\bigskip

1. It is easy to see that the recursion~\eq{eq:5} enumerates all planar arc structures on points
$x_i, x_{i + 1}, \dots, x_{i + k}$. In particular it implies that
\be
F_{i, i + k} \le \eps_{i, i + k} + F_{i + 1, k - 1}
\label{eq:103}
\ee for all $i$ and all odd~$k \ge 1$ and that
\be
F_{i, i + k} \le F_{i, i + \ell} + F_{i + \ell + 1, i + k}
\label{eq:100}
\ee
for all $i$ and $1 \le \ell < k$ with $k, \ell$ odd.  The latter property can be described as
\emph{subadditivity} of the functional~$F$: for two non-overlapping configurations of points $x_1 <
x_2 < \dots < x_{i + \ell}$ and $x_{i + \ell + 1} < x_{i + \ell + 2} < \dots < x_{i + k}$, the
value $F_{i, i + k}$ for the united configuration is not greater than the sum of the values $F_{i,
i + \ell}$ and $F_{i + \ell + 1, i + k}$ on the two partial configurations.

\bigskip

2. For the cost function $w(x_i, x_j) = \eps_{ij}$ of concave type, the free energy functional is
not only subadditive, but enjoys a stronger property: for all~$i$, odd $1 < \ell < k$ and even $j$
with $j \le \ell + 1$, $F$ verifies the inequality
\be
F_{i, i + k} + F_{i + j, i + \ell} \le F_{i, i + \ell} + F_{i + j, i + k}
\label{eq:102}
\ee
of which \eq{eq:100} is a particular case corresponding to~$j = \ell + 1$.  This property of~$F$ is
called \emph{submodularity}: note that it is similar to~\eq{eq:101} when $x_1 < x_2 < y_2 < y_1$.
It suffices to establish submodularity for $j = 2$ and~$\ell = k - 2$:
\be
F_{i, i + k} \le F_{i, i + k - 2} + F_{i + 2, i + k} - F_{i + 2, i + k - 2};
\label{eq:104}
\ee
the general case~\eq{eq:102} is recovered by induction.  Indeed, it was established
in~\cite{Delon.J:2011b} that $F$ satisfies a recursion
\begin{multline}
  F_{i, i + k} = \min \bigl[\eps_{i, i + k} + F_{i + 1,k - 1}; \\
  F_{i, i + k - 2} + F_{i + 2, i + k} - F_{i + 2, i + k - 2} \bigr]
\label{eq:6}
\end{multline}
that combines \eq{eq:103} and~\eq{eq:104}.  In other words, $F$ is the \emph{maximal} submodular
functional that satisfies also~\eq{eq:103}.

Thus, the function $F_{i,i+k}$ for concave--type potentials satisfies not only the standard
\emph{nonlocal} Eq.\eq{eq:5}, but also a \emph{local} Eq.\eq{eq:6}. For completeness a derivation
of~\eq{eq:6} taken from~\cite{Delon.J:2011b} is included in Appendix.

\section{The topological properties of the Random Interval Model}

The random interval model defined above has some interesting topological features. Namely, the
height diagram, $h$, which can be regarded as a quantitative characteristics of the ``nesting
degree'' of planar arcs, displays for the Gaussian distribution of intervals a topological
crossover from sequential pairing of monomers to essentially embedded (i.e. nested) one. Another
interesting behavior of $h$ is observed for the power--law (i.e. scale--free) distribution of
intervals, where the dependence of the height on the the exponent in the distribution has a
well--defined maximum.

\subsection{Numerical results}

\subsubsection{The truncated Gaussian distribution}

\begin{figure}[ht]
\epsfig{file=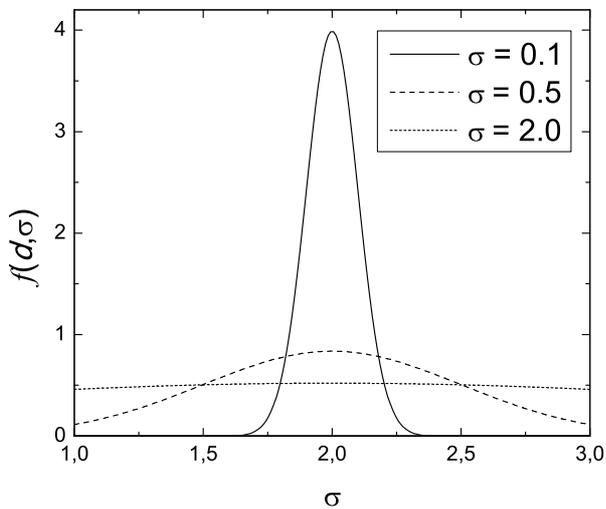, width=8cm}
\caption{Truncated Gaussian distribution $f(\sigma)$ of distances between nearest--neighboring
monomers, $\sigma=0.1;0.5;2.0$.}
\label{fig:3}
\end{figure}

Consider a random chain, in which the distances between nearest--neighboring monomers,
$d_i=|x_{i+1}-x_{i}|$, are distributed with the truncated Gaussian distribution:
\be
f(d,\sigma)=\left\{\begin{array}{ll} \disp
\frac{C}{\sqrt{2\pi}\sigma}e^{-\frac{(d-\mu)^{2}}{2\sigma^{2}}}, & \quad d_{\rm min}<d<d_{\rm max}
\\ 0,  & \quad \mbox{otherwise} \end{array}\right.
\label{eq:12}
\ee
where $C=2\left[{\rm erf}\left(\frac{d_{\rm max}-\mu}{\sqrt{2}\sigma}\right)+{\rm
erf}\left(\frac{\mu-d_{\rm min}}{\sqrt{2}\sigma}\right)\right]^{-1}$ is the constant determined by
the normalization condition $\int^{d_{\rm max}}_{d_{\rm min}}f(x,\sigma)\, dx=1$. To avoid any
possible misunderstandings, require all energies in \eq{eq:4} to be positive. Without te loss of
generality we can chose the following values of the parameters of the distribution function in
\eq{eq:12}: $\mu=2;\; d_{\rm min}=1;\; d_{\rm max}=3$. The distribution function \eq{eq:12} is
depicted in the \fig{fig:3} for different dispersions $\sigma$.

Our numerical analysis shows the existence of a crossover for random interval RNAs in topology of
monomer pairings (planar diagrams) from sequential to essentially nested one. The parameter which
controls this behavior is the dispersion $\sigma$ of the distribution $f(d,\sigma)$.

\begin{figure}[ht]
\epsfig{file=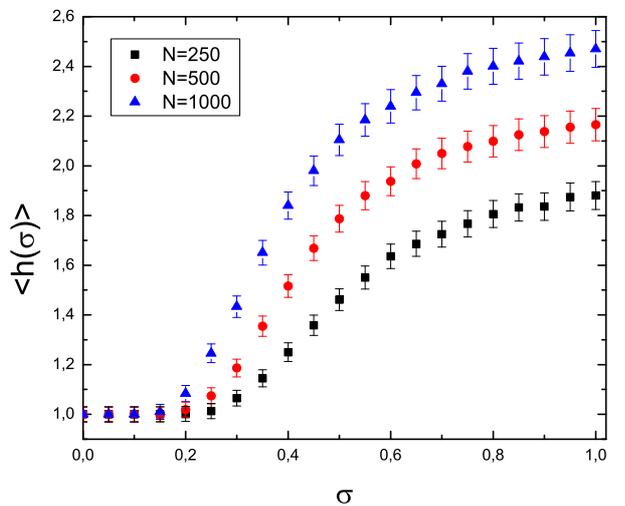, width=8cm}
\caption{Dependence of the average height, $\la h \ra$ on the control parameter $\sigma$ for
the Gaussian truncated distribution.}
\label{fig:4}
\end{figure}

For $\sigma<\sigma_{\rm cr}$, i.e. for essentially peaked distributions, the ground state of a
random RNA chain has a height equal to $1$. This means that only sequential pairs of nearest
neighboring monomers do form bonds. The value $\sigma_{\rm cr}$, at which the height diagram
exceeds $1$, we call the topological crossover point. The value $\sigma_{\rm cr}$ is computed for
finite chains and depends on its total length, $N$; when $N$ is increasing, the point of transition
shifts towards smaller values and, apparently, reaches zero when $N$ tends to infinity. The figure
\ref{fig:4} presents our numerical results for random interval chain with $N=250,500, 1000$
monomers.

Above the crossover point, i.e. for $\sigma>\sigma_{\rm cr}$ the height diagram monotonically
increases with $\sigma$ and reaches some averaged stationary value for the RIM with uniform
distribution of intervals ($\sigma\to \infty$). We prefer to use the term ``crossover'' instead of
``transition'' since we expect that it is not a true phase transition, the width of which shrinks
to zero in the thermodynamic limit.

\subsubsection{The power--law distribution}

The truncated Gaussian distribution considered above is good for testing the key features of the
RIM of RNA--like chains, however itself this distribution is rather artificial. It is much more
natural to consider the scale--free (power--law) distributions of distances between neighboring
monomers. In this case the intervals $d_i$ have the following probability density function:
\be
f(d,\gamma)=\frac{C}{1+d^{\gamma}}
\label{eq:16}
\ee
We consider all values $\gamma>0$ and truncate the distribution \eq{eq:16} outside the interval
$d_{\rm min}<d<d_{\rm max}$. The normalization constant $C\equiv C_{\gamma}(d_{\rm max}, d_{\rm
min})$ is
\be
\begin{array}{rll}
C(d_{\rm max}, d_{\rm min}) & = & \left[A_{\gamma}(d_{\rm max}) - A_{\gamma}(d_{\rm
min})\right]^{-1}; \medskip \\ A_{\gamma}(x) & = &
_2F_1\left(1,\gamma^{-1},1+\gamma^{-1},-x^{\gamma}\right)x
\end{array}
\label{eq:16a}
\ee
where $_2F_1(...)$ is the hypergeometric function. In what follows we take the following numerical
values: $d_{\rm min}=1;\; d_{\rm max}=20$. In contrast to the truncated Gaussian distribution, in
the truncated scale--free distribution the probability of very long distances between neighboring
monomers is not exponentially small.

\begin{figure}[ht]
\epsfig{file=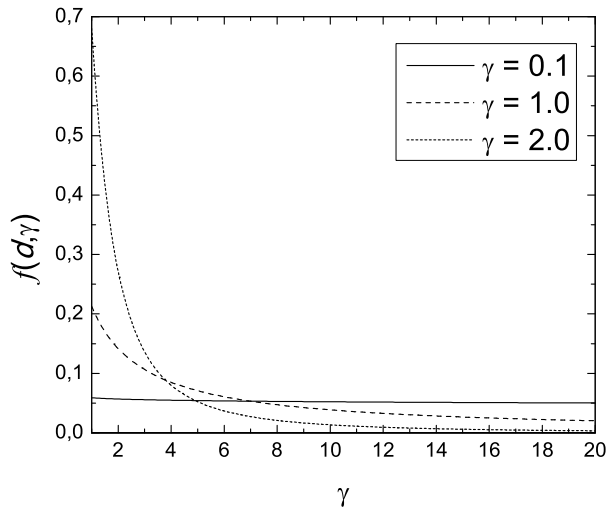, width=8cm}
\caption{Power--law distribution function $f(d,\gamma)$ of distances between nearest--neighboring
monomers, $\gamma=0.1;1.0;2.0$.}
\label{fig:5}
\end{figure}

\begin{figure}[ht]
\epsfig{file=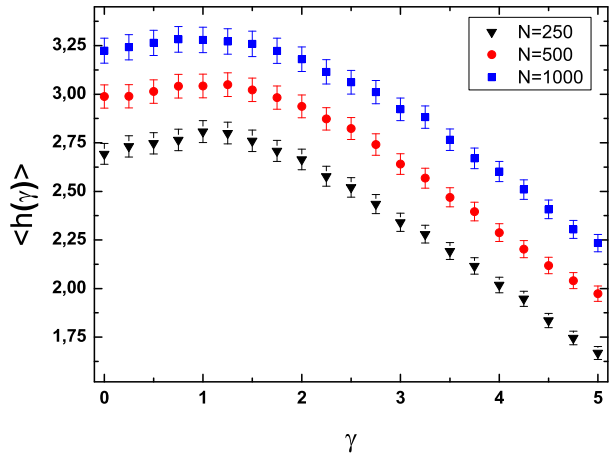, width=8cm}
\caption{Dependence of the height, $\la h \ra$ on the control parameter $\gamma$ for the truncated
power--law distribution.}
\label{fig:6}
\end{figure}

The presence of ``heavy tails'' in the distribution \eq{eq:16} affects the topology of the ground
state of the RNA RIM in a nontrivial way. Indeed, when $\gamma$ in \eq{eq:16} is increasing from
zero, the ``nesting degree'', $h$, behaves non-monotonically: at small $\gamma>0$ it increases up
to some maximal value (at $\gamma=1$) and then decreases, tending to 1 (for $\gamma \to \infty$) --
see the \fig{fig:6}.

It is worth to note that the presence of ``heavy tails'' in the distribution releases the creation
of nested configurations in an optimal pairing. For large values of $\gamma$ the height diagram
decreases which, as in the case of Gaussian distribution, corresponds to weakly random (practically
equidistant) RNAs with sequential optimal pairing.

\subsection{Analytic estimates}

The nesting in an optimal configuration of RIM is affected two complimentary factors. On one hand,
the nesting becomes favorable under some condition (explicitly written below) on lengths of three
sequential intervals $d_{i-1},\; d_{i},\; d_{i+1}$. On the other hand, the creation of a covering
arc between two distant monomers $i$ and $j$ could be favorable if below this arc all pairs of {\em
neighboring} monomers have formed bonds. Creation of a covering arc involves a global
reorganization of linked pairs in a RIM. To the contrary, the nesting discussed above, is the local
property of the RIM due to the special arrangement of sequential triples.

Let us focus on the nesting in an optimal configuration dealing with {\em local} properties of a
RIM. According to \eq{eq:102}--\eq{eq:104} the nested configuration of two arcs is favorable with
respect to the sequential pairing, if the following inequality for the values $\omega_{i-1,i+2},\;
\omega_{i-1,i},\; \omega_{i,i+1},\; \omega_{i+1,i+2}$ holds:
\be
\omega_{i-1,i+2}+\omega_{i,i+1}<\omega_{i-1,i}+\omega_{i+1,i+2}
\label{eq:13}
\ee
Taking into account that $\omega_{i,j}=u\ln|x_i-x_j|$, we can easily transform \eq{eq:13} into the
condition on three sequential intervals $d_{i-1},\; d_i,\; d_{i+1}$:
\be
\left\{\begin{array}{l} \disp d_{i-1}>d_{i} \medskip \\
\disp d_{i+1}>\frac{d_{i}(d_{i-1}+d_{i})}{d_{i-1}-d_{i}}
\end{array} \right.
\label{eq:14}
\ee
or in a more symmetric form
\be
d_i < \frac{d_{i - 1} + d_{i + 1}}2 \left(\sqrt{1 + \frac{4d_{i - 1}d_{i + 1}}{(d_{i - 1} + d_{i +
1})^2}} - 1\right).
\label{eq:14a}
\ee
It can be easily checked that \eqref{eq:14a} implies the first inequality~\eqref{eq:14}. Having the
distribution $f(d)$ (Gaussian, defined by \eq{eq:12}, or power--law, defined by \eq{eq:16})
truncated outside of the interval $[d_{\rm min}, d_{\rm max}]$, we can compute the probability $P$
that inequalities \eq{eq:14} hold. Since the intervals $d_{i-1},\; d_i,\; d_{i+1}$ are distributed
independently, the desired probability $P$ is determines by the integral
\begin{multline}
P = \int_{d_{\rm min}}^{d_{\rm max}} f(x)\, dx \int_{d_{\rm min}}^{d_{\rm max}} f(y)\, dy \\
\times \int_{d_{\rm min}}^{\frac{x + y}2\left(\sqrt{1 + \frac{4xy}{(x + y)^2}} - 1\right)} f(z)\,
dz,
\label{eq:15}
\end{multline}
where integration over $x$ corresponds to~$d_{i - 1}$, over $y$, to~$d_{i + 1}$, and over~$z$, to~$d_i$.

Equation \eq{eq:15} describes appearance of 1st level nesting ($h=2$). Moreover, it is present as a
multiplier in the probability of the 2nd level nesting ($h=3$). So, we can expect that numerical
curves for $h(\sigma)$ or $h(\gamma)$ have the same features as the function \eq{eq:15} for
distributions $f(d,\sigma)$ (Gaussian) and $f(d,\gamma)$ (power--law) respectively.

\subsubsection{Gaussian truncated distribution}

\begin{figure}[ht]
\epsfig{file=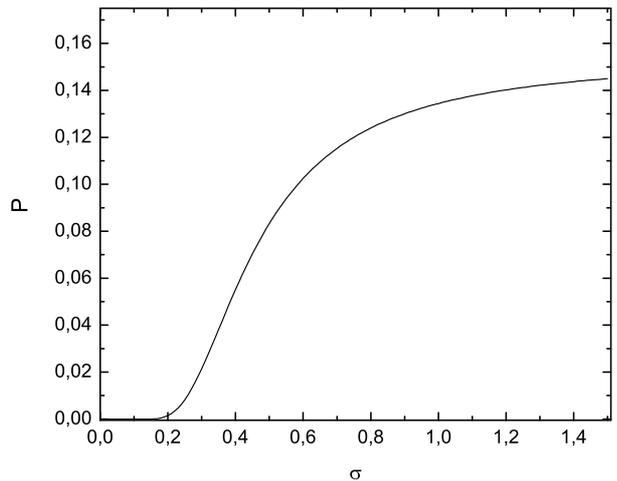, width=8cm}
\caption{Dependence of the probability $P$ (see \eq{eq:15}) on the control parameter $\sigma$
for the truncated Gaussian distribution.}
\label{fig:7}
\end{figure}

Substituting the truncated Gaussian distribution $f(d,\sigma)$ (see Eq. \eq{eq:12}) with the
parameters $\mu=2;\; d_{\rm min}=1;\; d_{\rm max}=3$ for $f(d)$ in Eq. \eq{eq:15}, we get the
function $P$ plotted in the \fig{fig:7}. Note that $P(\sigma)$ repeats the profile of $\la
h(\sigma)\ra$ displayed in the \fig{fig:4} for the average height of the arc diagram. However our
analytic approach does not take into account the slight dependence of the transition point on the
polymer length since this effect has ``global'' property and is beyond the precision of our method.
It should be also emphasized that the appearance of the 2nd--level nesting (i.e. of the diagrams
with the heights $h>2$) deals exclusively with global reorganization of pairing in the RIM. Indeed,
in order to have the 2nd level nesting, the condition \eq{eq:14} should be valid for the intervals
$d_{i-2},\; d^{(1)},\; d_{i+2}$, where we substitute for the middle interval $d^{(1)}$ the
combination of neighboring triples, $d_{i-1}+ d_{i}+ d_{i+1}$, which itself is nested. The minimal
value for the middle interval $d^{(1)}$, as it follows from \eq{eq:14}, is $d^{(1)}=2(\sqrt{2}+1)
d_{\rm min} +d_{\rm min}$. For the parameters of our distribution, we can conclude, that
$d^{(1)}>d_{\rm max}$, what contradicts with the definition of the model. It means that all the
configurations with the $h>2$ have at least one long ``global'' arc.

\subsubsection{Power--law truncated distribution}

The same analysis can be performed for the RIM with the power--law distribution $f(d,\gamma)$ (see
Eq. \eq{eq:16}). The presence of nested structures in an optimal pairing is determined by the
function $P$ \eq{eq:15}, which now depends on the parameter $\gamma$ in the distribution
\eq{eq:16}. We see that the function $P(\gamma)$ has the maximum at the point $\gamma=1$. At
$\gamma\gg 1$ the probability $P$ tends to zero. Contrary to the truncated Gaussian distribution,
the 2nd level nesting is allowed since $d^{(1)}<d_{\rm max}$, however the 3rd level nesting is
forbidden, because $d^{(2)}=2(\sqrt{2}+1)d^{(1)} + d^{(1)}>d_{\rm max}$. So, in the configurations
with $h>3$ the nesting is again due to ``global'' factors.

\begin{figure}[ht]
\epsfig{file=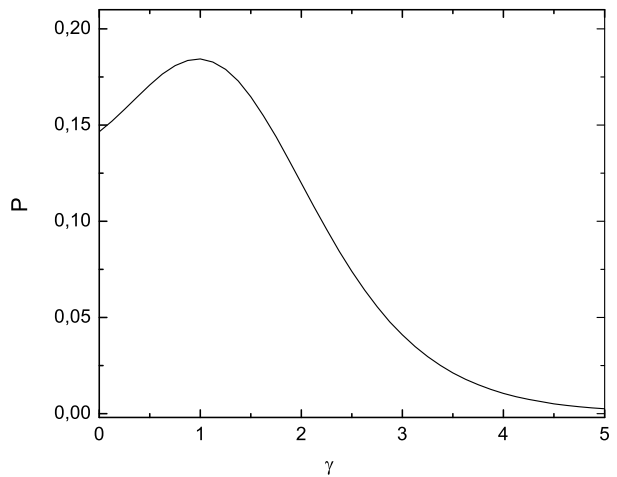, width=8cm}
\caption{Dependence of the probability $P$ (see \eq{eq:15}) on the control parameter $\gamma$
for the truncated power--law distribution.}
\label{fig:8}
\end{figure}

\section{Conclusion}

In this paper we have proposed a new model of a heteropolymer chain with RNA--type topology of
secondary structure and quenched random distribution of intervals between neighboring monomers. For
quantitative analysis of the Random Interval Model (RIM), we have investigated the statistical
behavior of ``height diagrams'' as a function of the control parameter in the distribution function
of intervals.

We have shown that for truncated Gaussian distribution $f(d,\sigma)$ of intervals (see Eq.
\eq{eq:12}), the height diagram exhibits a topological transition in pairing of monomers from
sequential to essentially nested one. The parameter which controls this behavior is the dispersion,
$\sigma$, of the distribution $f(d,\sigma)$.

In contrast to the truncated Gaussian distribution, for the truncated scale--free distribution
$f(d,\gamma)$ (see Eq. \eq{eq:16}) the probability of very long distances between neighboring
monomers is not exponentially small. The presence of such ``heavy tails'', or, in other words, of
the ``intermittent behavior'' (i.e. very long tails mixed with very short ones) nontrivially
affects the topology of the ground state of the RNA Random Interval Model. Indeed, when $\gamma$ in
\eq{eq:16} is increasing from zero, the ``nesting degree'', $h$, behaves non-monotonically: at
small $\gamma>0$ it increases up to some maximal value (at $\gamma=1$) and then decreases, tending
to 1 (for $\gamma \to \infty$).

The important result deserving attention, concerns the possibility to pass from the {\em nonlocal}
recursion relation for the ground state free energy \eq{eq:5} to the local recursion relation
\eq{eq:6} if and only if the interaction energy between paired monomers, $\eps_{i,j}$, is a concave
function of distance. So, for any potential (even random) of concave form, the equation \eq{eq:6}
(and, hence, Eq. \eq{eq:1}) can be essentially simplified resulting in shortening the computational
time if these equations are implemented for numeric analysis of secondary structures of polymer
chain with RNA--type architecture.

The final remark concerns the possible interplay between optimization problems and some particular
results of the Random Matrix Theory (RMT) for RNA folding, addressed in \cite{orland1,orland2}. Let
us recall that our basic result relays on the theorem which proves that optimal pairings on the
line with the concave transport function are non-intersecting (i.e. planar) -- see, for example,
the \fig{fig:example}. Being formulated in RMT terms, this means that optimization leads to the
extraction of a special subclass of planar diagrams in the large--$N$ random matrix ensemble,
namely, the so-called {\em rainbow diagrams} -- see, for example, \cite{gudowska}. To this end it
would be interesting to formulate our Random Interval Model as a matrix model for finite $N$ in
order to check how the optimization algorithms allow extract planar diagrams of special topology in
matrix models.

We are grateful to V.~Avetisov, K.~Khanin, S.~Majumdar and M.~Tamm for various discussions of the
problem. S.K.N and O.V.V. are partially maintained by the European Network ERASysBio+\#66 "GRAPPLE"
and by the ANR grant 2011-BS04-013-01 ``WALKMAT.'' A.S. acknowledges the IRSES project 269139
DCP-PhysBio, the RFBR grant 11-01-93106 CNRSL\_a, and the RF government grant 11.G34.31.0073.

\appendix

\section{Derivation of Eq.\eq{eq:6}}

Suppose $X = \{x_i\}_{1\le i\le 2n}$ with $x_1 < x_2 < \dots < x_{2n}$ and $X' = \{x'_{i'}\}_{1\le
i'\le 2n'}$ with $x'_1 < x'_2 < \dots < x'_{2n'}$ are two sets such that $x_{2n} < x'_1$, i.e.,
$X'$ lies to the right of~$X$.

We will refer to minimum--weight perfect matchings on $X$ and~$X'$,
i.e., planar (nonintersecting) sets of $n$ (resp.\ $n'$)
arcs connecting the points such that the sum of their weights, which are
given by a cost function $w(\cdot, \cdot)$ of concave type, is minimal, as \emph{partial matchings}
and to the minimum--weight perfect matching on~$X \cup X'$ as \emph{joint matching}.

Call an arc $(x_i, x_j)$ in a nested matching \emph{exposed} if there is no arc $(x_{i'}, x_{j'})$
with $x_i, x_j$ contained between $x_{i'}$ and~$x_{j'}$. We call all other arcs in a nested
matching non-exposed or \emph{hidden}. Intuitively, exposed arcs are those visible ``from above''
and hidden arcs are those covered with exposed ones.

We first show, following~\cite{Delon.J:2011b}, that whenever an arc $(x_i, x_j)$ is hidden in the
partial matching on~$X$, it belongs to the joint optimal matching and is hidden there too. By
contradiction, assume that some of hidden arcs in the partial matching on~$X$ do not belong to the
joint matching. Then there will be at least one exposed arc $(x_\ell, x_r)$ in the partial matching
on~$X$ such that some points $x_i$ with $x_\ell < x_i < x_r$ are connected in the joint matching to
points outside $(x_\ell, x_r)$. %

Denote all the points in the segment $[x_\ell, x_r]$ that are connected in the joint matching to
points on the left of~$x_\ell$ by $z_1 < z_2 < \dots < z_k$; denote the opposite endpoints of the
corresponding arcs by $y_1 > y_2 > \dots > y_k$, where the inequalities follow from the fact that
the joint matching is nested. Likewise denote those points from~$[x_\ell, x_r]$ that are connected
in the joint matching to points on the right of~$x_r$ by $z'_1 > z'_2 > \dots > z'_{k'}$ and their
counterparts in the joint matching by $y'_1 < y'_2 < \dots < y'_{k'}$. Observe that although $k$ or
$k'$ may be zero, the number $k + k'$ must be positive and even.

Consider now a matching on the segment $[x_\ell, x_r]$ that consists of the following arcs: those
arcs of the joint matching whose both ends belong to~$[x_\ell, x_r]$; the arcs $(z_1, z_2)$, \dots,
$(z_{2\kappa - 1}, z_{2\kappa})$, where \footnote{$\lfloor \xi\rfloor$ is the largest integer $n$
such that $n\le \xi$.} $\kappa = \lfloor k/2\rfloor$; the arcs $(z'_2, z'_1)$, \dots,
$(z'_{2\kappa'}, z'_{2\kappa' - 1})$, where $\kappa' = \lfloor k'/2\rfloor$; and $(z_k, z'_{k'})$
if both $k$ and~$k'$ are odd. Denote by~$W'$ the weight of this matching. It cannot be smaller than
the weight~$W'_0$ of the restriction of the optimal partial matching on~$X$ to $[x_\ell, x_r]$. For
the total weight~$W$ of the joint matching on $X \cup X'$ we thus have
\begin{equation}
  \label{eq:01}
  W \ge W - W' + W'_0.
\end{equation}
We now show that by a suitable sequence of uncrossings the right--hand side here can be further
reduced to a matching whose weight is strictly less than~$W$.

The arcs $(z_1, y_1)$ and~$(x_\ell, x_r)$ are crossing, so that $w(y_1, z_1) + w(x_\ell, x_r) >
w(y_1, x_\ell) + w(z_1, x_r)$. Uncrossing these arcs strictly reduces the right-hand side
of~\eqref{eq:01}:
\begin{multline*}
  W > W - W' + W'_0 \\- w(y_1, z_1) - w(x_\ell, x_r) + w(y_1, x_\ell)
  + w(z_1, x_r).
\end{multline*}
\begin{widetext}
  Now the arcs $(y_2, z_2)$ and $(z_1, x_r)$ are crossing, so $w(y_2,
  z_2) + w(z_1, x_r) - w(z_1, z_2) > w(y_2, x_r)$ and therefore
  \begin{displaymath}
    W > W - W' + W'_0 - w(y_1, z_1) - w(y_2, z_2) - w(x_\ell, x_r)
    + w(y_1, x_\ell) + w(z_1, z_2) + w(y_2, x_r).
  \end{displaymath}
\end{widetext}
Repeating this step $\kappa = \lfloor k/2\rfloor$ times gives the inequality
\begin{multline*}
  W > W - W' + W'_0 - w(x_\ell, x_r) - \sum_{1\le i\le 2\kappa} w(y_i, z_i) \\
  + \sum_{1\le i\le \kappa}\!\! w(z_{2i - 1}, z_{2i}) + \sum_{1\le
    i\le \kappa}\!\! w(y_{2i - 1}, y_{2i - 2}) + w(y_{2\kappa}, x_r),
\end{multline*}
where in the rightmost sum $y_0$ is defined to be~$x_\ell$. Note that at this stage all arcs coming
to points $z_1, z_2, \dots$ from outside $[x_\ell, x_r]$ are eliminated from the matching, except
possibly $(y_k, z_k)$ if $k$ is odd.

It is now clear by symmetry that a similar reduction step can be performed on arcs going from
$z'_1, z'_2, \dots$ to the right.

Finally if $k$ and $k'$ are odd, we uncross the pair of arcs $(y_k, x_k)$ and $(y_{k - 1}, y'_{k' -
1}$ and finally the pair $(z_k, y'_{k' - 1})$ and $(z'_{k'}, y'_{k'})$.

\begin{widetext}
  The final estimate for~$W$ has the form
  \begin{multline}
    \label{eq:02}
    W > W - W' + W'_0 - w(x_\ell, x_r) - \sum_{1\le i\le k} w(y_i, z_i)
    - \sum_{1\le i'\le k'} w(z'_{i'}, y'_{i'}) \\
    + \sum_{1\le i\le \kappa} w(z_{2i - 1}, z_{2i})
    + \sum_{1\le i'\le \kappa'} w(z'_{2i'}, z'_{2i' - 1})
    + w(z_k, z'_{k'})\cdot [\text{$k$, $k'$ are odd}] \\
    + \sum_{1\le i\le\kappa} w(y_{2i - 1}, y_{2i - 2})
    + \sum_{1\le i'\le\kappa'} w(y'_{2i' - 2}, y'_{2i' - 1})
    + w(y_k, y'_{k'})\cdot[\text{$k$, $k'$ are even}],
  \end{multline}
  where notation such as [$k$, $k'$ are odd] means~$1$ if $k$, $k'$
  are odd and~$0$ otherwise.\\
\end{widetext}

The right--hand side of~\eqref{eq:02} contains four groups of terms: first,
\begin{displaymath}
  W - \sum_{1\le i\le k} w(y_i, z_i) - \sum_{1\le i'\le k'} w(z'_{i'}, y'_{i'}),
\end{displaymath}
corresponding to the joint matching without the arcs connecting points
inside $[x_\ell, x_r]$ to points outside this segment; second,
\begin{multline*}
  W' - \sum_{1\le i\le \kappa} w(z_{2i - 1}, z_{2i})
  - \sum_{1\le i'\le\kappa'} w(z'_{2i'}, z'_{2i' - 1}) \\
  - w(z_k, z'_{k'})\cdot[\text{$k$, $k'$ are odd}],
\end{multline*}
which comes with a negative sign and corresponds to the arcs of the joint matching with both ends
inside~$[x_\ell, x_r]$, and cancels them from the total; third, $W'_0 - w(x_\ell, x_r)$, with
positive sign, which corresponds to the hidden arcs of the partial matching on~$X$ inside the
exposed arc $(x_\ell, x_r)$, not including the latter; and finally the terms in the last line
of~\eqref{eq:02}, corresponding to the arcs matching $x_\ell$, $x_r$, and points $y_1, \dots, y_k,
y'_1, \dots, y'_{k'}$, i.e., those points outside $[x_\ell, x_r]$ that were connected in the joint
matching to points inside this segment.

Gathering together contributions of these four groups of terms, we observe that all negative terms
cancel out and what is left corresponds to a perfect matching with a weight strictly smaller
than~$W$, in which all arcs hidden by $(x_\ell, x_r)$ in the partial matching on~$X$ are restored.
There may still be some crossings caused by terms of the fourth group and \emph{not} involving the
hidden arcs in $[x_\ell, x_r]$; uncrossing them if necessary gives a nested perfect matching whose
weight is strictly less than that of the joint matching. This contradicts the assumption that the
latter is the minimum--weight matching on~$X\cup X'$. Therefore all hidden arcs in the partial
matching on~$X$ (and, by symmetry, those in the partial matching on~$X'$) belong to the joint
matching.

Now let $i$,~$j$ be indices of opposite parity and such that $i < j$, and define $W_{i,j}$ to be
the weight of the minimum-weight perfect matching on the $j - i + 1$ points~$x_i < x_{i + 1} <
\dots < x_j$. We can now show, following~\cite{Delon.J:2011b}, that for all indices $i$,~$j$ of
opposite parity with $1 \le i < j \le 2n$, weights~$W_{i, j}$ satisfy the recursion
\begin{multline}
  \label{eq:03}
  W_{i, j} = \min\, \bigl[w(x_i, x_j) + W_{i + 1, j - 1};\\
  W_{i, j - 2} + W_{i + 2, j} - W_{i + 2, j - 2}\bigr]
\end{multline}
with ``initial conditions''
\begin{equation}
  \label{eq:04}
  W_{i, i - 1} = 0,\quad W_{i + 2, i - 1} = -w(x_i, x_{i + 1}).
\end{equation}

For simplicity we will refer to the minimum-weight perfect matching on points $x_r < x_{r + 1} <
\dots < x_s$ as the ``matching~$W_{r, s}$.'' Consider first the matching that consists of the
arc~$(x_i, x_j)$ and all arcs of the matching~$W_{i + 1, j - 1}$, and observe that by optimality
the latter its weight $w(x_i, x_j) + W_{i + 1, j - 1}$ is minimal among all matchings that
contain~$(x_i, x_j)$.

We now examine the meaning of the expression $W_{i, j - 2} + W_{i + 2, j} - W_{i + 2, j - 2}$.
Denote the point connected in the matching~$W_{i, j - 2}$ to~$x_i$ by~$x_k$ and the point connected
to~$x_{i + 1}$ by~$x_\ell$. It is easy to see that the pairs of indices $i, k$ and $i + 1, \ell$
both have opposite parity. Assume first that
\begin{equation}
  \label{eq:05}
  x_{i + 1} < x_\ell < x_k \le x_{j - 2}.
\end{equation}

Observing that hidden arcs in partial matchings on the sets $X = \{x_i, x_{i + 1}\}$ and $X' =
\{x_{i + 2}, \dots, x_{j - 2}\}$ are preserved, and taking into account parity of $k$ and~$\ell$,
we see that $x_k$ and~$x_\ell$ (as well as their neighbors $x_{k + 1}$ and~$x_{\ell - 1}$ if they
are contained in~$[x_{i + 2}, x_{j - 2}]$) belong to exposed arcs of the matching~$W_{i + 2, j -
2}$. Thus the matching~$W_{i, j - 2}$ has the following structure:

\begin{figure}[ht]
\epsfig{file=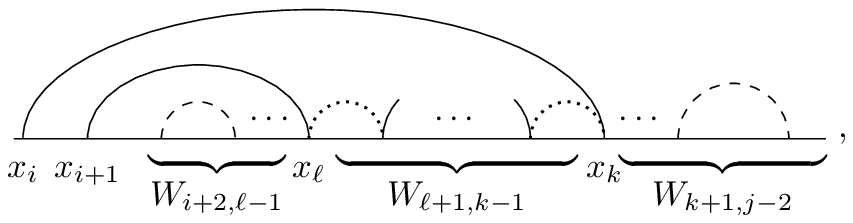, width=8cm}
\label{fig:arcs_fig}
\end{figure}

where dashed (resp., dotted) arcs correspond to those exposed arcs of the matching~$W_{i + 2, j -
2}$ that belong (resp., do not belong) to~$W_{i, j - 2}$.

Since points $x_{\ell - 1}$ and~$x_{k + 1}$ belong to exposed arcs in the matching~$W_{i + 2, j -
2}$, the (possibly empty) parts of this matching that correspond to points $x_{i + 2} < \dots <
x_{\ell - 1}$ and $x_{k + 1} < \dots < x_{j - 2}$ coincide with the (possibly empty) matchings
$W_{i + 2, \ell - 1}$ and~$W_{k + 1, j - 2}$. For the same reason the (possibly empty) part of the
matching~$W_{i, j
  - 2}$ supported on $x_{\ell + 1} < \dots < x_{k - 1}$ coincides
with~$W_{\ell + 1, k -1}$. %
Therefore
\begin{multline}
  \label{eq:06}
  W_{i, j - 2} = w(x_i, x_k) + w(x_{i + 1}, x_\ell)\\ + W_{i + 2, \ell - 1}
  + W_{\ell + 1, k - 1} + W_{k + 1, j - 2}.
\end{multline}
Taking into account~\eqref{eq:04}, observe that in the case $k = i + 1$ and $\ell = i$, which was
left out in~\eqref{eq:05}, this expression still gives the correct formula $W_{i, j - 2} = w(x_i,
x_{i + 1}) + W_{i + 2, j - 2}$.

Now assume that in the matching $W_{i + 1, j}$ the point $x_j$ is connected to $x_{\ell'}$ and the
point~$x_{j - 1}$ to~$x_{k'}$. A similar argument gives
\begin{multline}
  \label{eq:07}
  W_{i + 2, j} = W_{i + 2, \ell' - 1} + W_{\ell' + 1, k' - 1} + W_{k' + 1,
    j - 2} \\+ w(x_{\ell'}, x_j) + w(x_{k'}, x_{j - 1});
\end{multline}
in particular, if $\ell' = j - 1$ and $k' = j$, then $W_{i + 2, j} =
W_{i + 2, j - 2} + w(x_{j - 1}, x_j)$. %

Suppose that $x_k < x_{\ell'}$. Taking into account that $x_k$, $x_{k + 1}$, $x_{\ell' - 1}$,
and~$x_{\ell'}$ all belong to exposed arcs in~$W_{i + 2, j - 2}$, we can write
\begin{equation}
  \label{eq:08}
  \begin{gathered}
    W_{k + 1, j - 2} = W_{k + 1, \ell' - 1} + W_{\ell', j - 2},\\
    W_{i + 2, \ell' - 1} = W_{i + 2, k} + W_{k + 1, \ell' - 1}
  \end{gathered}
\end{equation}
and
\begin{equation}
  \label{eq:09}
  W_{i + 2, j - 2} = W_{i + 2, k} + W_{k + 1, \ell' - 1} + W_{\ell', j - 2}.
\end{equation}
Substituting~\eqref{eq:08} into \eqref{eq:06} and~\eqref{eq:07} and
taking into account~\eqref{eq:09}, we obtain
\begin{multline*}
  W_{i, j - 2} + W_{i + 2, j} - W_{i + 2, j - 2} = w(x_i, x_k) +
  w(x_{i + 1}, x_\ell) \\+ W_{i + 2, \ell - 1} + W_{\ell + 1,
    k - 1}
  + W_{k + 1, \ell' - 1} \\+ w(x_{\ell'}, x_j) + W_{\ell' + 1, k' - 1} +
  w(x_{k'}, x_{j - 1}) + W_{k' + 1, j - 2}.
\end{multline*}
The right-hand side of this expression corresponds to a matching that coincides with $W_{i, j - 2}$
on $[x_i, x_k]$, with $W_{i + 2, j - 2}$ on~$[x_{k + 1}, x_{\ell' - 1}]$, and with~$W_{i + 1, j}$
on $[x_{\ell'}, x_j]$. By optimality, this matching cannot be improved on any of these three
segments and is therefore optimal among all matchings in which $x_i$ and~$x_j$ belong to different
exposed arcs.

It follows that under the assumption that $x_k < x_{\ell'}$ the expression in the right-hand side
of~\eqref{eq:03} gives the minimum weight of all matchings on $x_i < x_{i + 1} < \dots < x_j$.
Moreover, the only possible candidates for the optimal matching are those constructed above: one
that corresponds to $w(x_i, x_j) + W_{i + 1, j - 1}$ and one given by the right-hand side of the
latter formula.

It remains to consider the case $x_k \ge x_{\ell'}$. Since $x_k \neq x_{\ell'}$ for parity reasons,
it follows that $x_k > x_{\ell'}$; now a construction similar to the above yields a matching which
corresponds to $W_{i, j - 2} + W_{i + 2, j} - W_{i + 2, j - 2}$ and in which the arcs $(x_i, x_k)$
and~$(x_{\ell'}, x_j)$ are crossed. Uncrossing them leads to a matching with strictly smaller
weight, which contains the arc~$(x_i, x_j)$ and therefore cannot be better than $w(x_i, x_j) + W_{i
+ 1, j - 1}$. This means that~\eqref{eq:03} holds in this case too with $W_{i, j} = w(x_i, x_j) +
W_{i + 1, j - 1}$.

\end{document}